\documentclass[iop]{emulateapj}

\newcommand{\Msun}{~M_\odot}
\newcommand{\msun}{M_\odot}

\newcommand{\ccm}{\rm ~cm^{-3}}
\newcommand{\kms}{\rm ~km~s^{-1}}

\newcommand{\wl}{\lambda}
\newcommand{\wll}{\lambda \lambda}

\begin{document}

\title{Reconciling the infrared catastrophe and observations of SN 2011fe}
\author{
Claes Fransson\altaffilmark{1},
Anders Jerkstrand\altaffilmark{2}
}
\altaffiltext{1}{Department of Astronomy, Oskar Klein Centre, Stockholm University, AlbaNova, 
SE--106~91 Stockholm, Sweden}
\altaffiltext{2}{Astrophysics Research Centre, School of Mathematics and Physics, Queen's University Belfast, Belfast, BT7 1NN, UK .}

\begin{abstract}
The observational effects of the 'Infrared Catastrophe'  are discussed in view of the very late observations of the Type Ia SN 2011fe. 
Our model spectra at 1000d take non-local radiative transfer into account, and find that this has a crucial impact on the spectral formation. 
Although rapid cooling of the ejecta to a few 100 K occurs also in these models, the late-time optical/NIR flux is brighter by 1-2 magnitudes due to redistribution of UV emissivity, resulting from non-thermal excitation and ionization. This effect brings models into  better agreement with late-time observations of SN 2011fe and other Type Ia supernovae, and offers a solution to the long standing discrepancy between models and observations. 
The models show that spectral formation shifts from \ion{Fe}{2} and \ion{Fe}{3} at 300d to \ion{Fe}{1} at 1000d, which explains the apparent wavelength shifts seen in SN2011fe. We discuss effects of time dependence and energy input from ${}^{57}$Co, finding both to be important at 1000d.

\end{abstract}


\section{INTRODUCTION}
\label{sec-introd}
The infrared catastrophe (IRC) of Type Ia supernovae (SNe) was predicted already in 1980  by \cite{Axelrod1980}.   It occurs at $\sim$500 days as a result of the change in cooling of the ejecta from transitions in the optical to fine-structure transitions in the mid- and far-infrared. Up to $\sim$500
days  cooling is dominated by optical/NIR [\ion{Fe}{2}{-\small III}] lines with excitation temperatures $\sim (1-3)\times10^4$ K. As radioactive energy input and temperature decrease, at $\sim 2000$ K the cooling is replaced by  mid-IR fine-structure lines with excitation temperatures $\sim 500$ K. 
Because the cooling is insensitive to temperature between $\sim$500-2000 K, the temperature drops dramatically until the Boltzmann factors for the fine structure lines become effective and the temperature stabilises at a few 100 K. At this point the thermal emission has moved completely to the mid/far-infrared.

\cite{Fransson1996} renewed this calculation,  
but found a strong discrepancy with the late observations of SN 1972E. 
Later studies \citep[e.g.,][]{Sollerman2004,Leloudas2009} confirmed this and did not find much evidence for the IRC from either photometry or spectra. 
The modeling in these papers was based on the  \citet[][b]{Kozma1998a} code, here referred  to as the KF98 code. 
  The model light curves predicted large drops in most photometric bands caused by the IRC, in disagreement with observations.
The redistribution of radiation caused by the multiple scattering and fluorescence could, however, not be treated.

The nearby Type Ia SNe 2011fe  and 2014J offer unique possibilities to study Ia SNe at late times with photometry  
and spectra up to $\sim 1000$ days. 
From photometry up to $\sim 950$ days \cite{Kerzendorf2014} found  a decline in the optical broadly consistent with that expected for ${}^{56}$Co decay,  with no signs of an IRC. 

In \citet[][in the following T2015]{Taubenberger2015} a spectrum taken at 1034 days was discussed with tentative line identifications. 
 From a comparison with a $\sim 300$ day spectrum it was clear that the spectrum  at this epoch was still dominated by Fe lines,  but it was not clear if these were from \ion{Fe}{1} or \ion{Fe}{2}. This is important as a direct  probe of the physical conditions of the ejecta.
 \cite{Graham2015} present a similar discussion based on a spectrum 981 days after explosion. These late epochs also offer a possibility to test  explosion models from their nucleosynthesis, in particular the formation of isotopes
such as ${}^{57}$Ni,  $^{55}$Fe and ${}^{44}$Ti   
\citep[e.g.][]{Ropke2012}. From photometry of SN 2012cg at $\sim 3$ years \cite{Graur2015} claim evidence for power input by ${}^{57}$Co, although the absence of spectral modeling makes the bolometric correction uncertain.

In this Letter we discuss a solution of the apparent contradiction between models of the IRC and observations, based on detailed spectral synthesis of realistic explosion models. 

\section{models}
\label{sec-mod}

To calculate spectra and light curves 
we use two different codes. One is based on KF98, updated with extended atomic data. This code includes time dependence and calculates the ionization and temperature of the ejecta, including non-thermal excitations, and {\it local} scatterings treated with the Sobolev approximation. This gives an accurate calculation of the temperature and ionization, but does not take into account the multiple scatterings which redistribute UV radiation into optical radiation.  As  discussed in \cite{Jerkstrand2011}, this is  important in the phase when non-thermal processes dominate the ionization and excitation.

 To treat this we use the code by \cite{Jerkstrand2011} in its latest version \citep{Jerkstrand2015}, here referred to as the JFK11+ code. This  treats the combined NLTE and radiative transfer problem for all relevant ions from hydrogen to nickel, but does not include time dependence. These two approaches are therefore complementary.

 As input explosion models we study the 1D W7 model by \cite{Nomoto1984}  and the N100 3D delayed detonation model by \cite{Seitenzahl2013}, which we make a 1D version of. For W7 we use the composition and density structure computed by \citet[][their Fig. 14]{Iwamoto1999}. The ejecta are zoned into 30 (KF98 calculation) and 180 (JFK11+) shells, with the larger number needed to resolve
the radiative transfer in the JFK11+ models.  

For N100 we
map the 3D model to a 1D version with the following recipe,  which minimises the microscopic mixing between the major burning zones, in contrast to a straight spherical average at the different radii:  For each cell in the 3D model the two most abundant elements are identified. This produces a set of 8 discrete composition classes. For each of these classes  the composition is  averaged  by mass over all cells belonging to
that class. The ejecta are then divided into shells in velocity up to $25,000 \kms$, with 40 shells for KF98 and 240 shells for JFK11+. For each radial shell the two composition classes with the highest mass within that velocity range in the 3D model are identified, and the shell is split into two equal-density subshells with corresponding compositions and relative masses  (more massive innermost). Compared to the original N100 model it conserves the total elemental masses at $\la 10\%$, with exception of ${}^{58}$Ni, with a total mass $0.099 \Msun$, compared to $0.069 \Msun$ in the original model. This is a consequence of using only two composition zones per radial shell.

Radioactive input (gamma-rays, leptons and X-rays) from ${}^{56}$Ni, ${}^{57}$Ni and ${}^{44}$Ti and their daughter isotopes are included \citep[e.g.,][]{Seitenzahl2009}. 
At 1000 days ${}^{56}$Ni and ${}^{57}$Ni contribute roughly equal to the deposition, depending on the exact ${}^{57}$Ni  mass. The ${}^{56}$Ni mass is $ 0.59 \  \msun$ in W7 and $ 0.60 \  \msun$ in N100, within the observed range for SN 2011fe, $ 0.5-0.6 \  \msun$ \citep{Pereira2013}. There is no independent observational constraint on the ${}^{57}$Ni mass. The delayed detonation models by 
\cite{Seitenzahl2013} have $(1.1-3.4)\times 10^{-2} \Msun$, with N100 at $1.8\times 10^{-2} \Msun$,  
while  W7 has $2.4 \times 10^{-2} \Msun$ \citep{Iwamoto1999}. 

In comparisons with the observations we have used a distance of 6.4 Mpc to M 101 \citep{Shappee2011}, and $E_{\rm B-V} = 0$ \citep{Nugent2011}.

\section{Results}
\label{sec-res}

\begin{figure}[!t]
\begin{center}
\resizebox{\hsize}{!}{\includegraphics[angle=0]{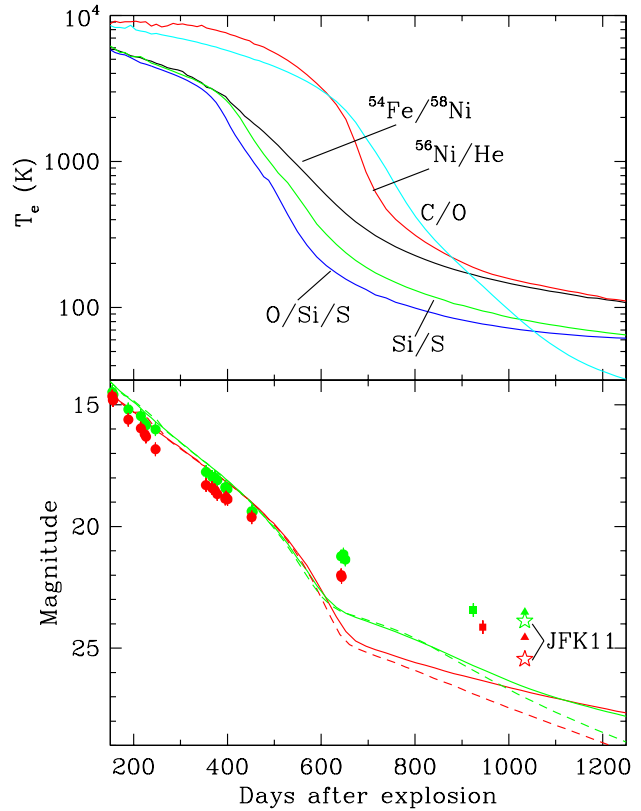}}
\caption{Upper panel: Temperature evolution of selected zones in the ejecta of the W7 model calculated with the KF98 code. Note the dramatic drop at $400-800$ days indicating  the transition from cooling by optical lines to cooling by mid-IR lines. The mean velocities of the zones are 3000 $\kms$ (${}^{54}$Fe/${}^{58}$Ni), 6800 $\kms$ (${}^{56}$Ni/He), 10,900 $\kms$ (Si/S), 13,000 $\kms$ (O/Si) and 15,000 $\kms$ (C/O). 
Lower panel: Synthetic light curves in the  V (green) and  R (red)  bands, for the same model together with observations in B and V at 0-652d \cite[circles,][]{Tsvetkov2013},  g and r at 930d  \cite[squares,][]{Kerzendorf2014}, and g and r at 1034d (triangles, T2015).  The open stars at 1034d are synthetic g and r magnitudes for the JFK11+ W7 model. The dashed lines show the V and R band light curves for a KF98 model in steady state, showing time-dependence to become important after 800-900d.}
\label{fig1b}
\end{center}
\end{figure}

In the upper panel of Fig. \ref{fig1b} we show the temperature  of typical composition zones of the W7 ejecta as function of time for the KF98 model. The temperature in all cases shows a drop from $(5 -10)\times 10^3$ K to $\la 100$ K. 
The epoch when this  
occurs differs from 400 to 800 days, depending on the composition. The C/O zone is cooling  more slowly than the others, but reaches the lowest temperatures after 1000 days.  This is a combination of low density, higher ionization and low cooling efficiency of the abundant ions.  The cooling is here dominated by [\ion{C}{2}] 157.7$\mu$m and [\ion{O}{3}] 88.34$\mu$m (excitation temperatures 91 K and 164 K, respectively), while the Si and Fe rich zones cool by fine-structure lines of higher energy, 410 K and 553 K for [\ion{Si}{2}] 34.81$\mu$m  and [\ion{Fe}{2}] 25.99$\mu$m, respectively. The cooling of these zones  therefore drops  at higher temperature compared to the C/O zone.

In the lower panel of Fig. \ref{fig1b} we show the V and R band light curves of the W7 model with the KF98 code. The IRC is here seen  as the drop at $\sim 500$ days in both bands.  A comparison with the magnitudes of \cite{Tsvetkov2013} and  \cite{Kerzendorf2014} of SN 2011fe clearly shows the discrepancy
after the IRC. 

\begin{figure}[!h]
\begin{center}
\resizebox{\hsize}{!}{\includegraphics{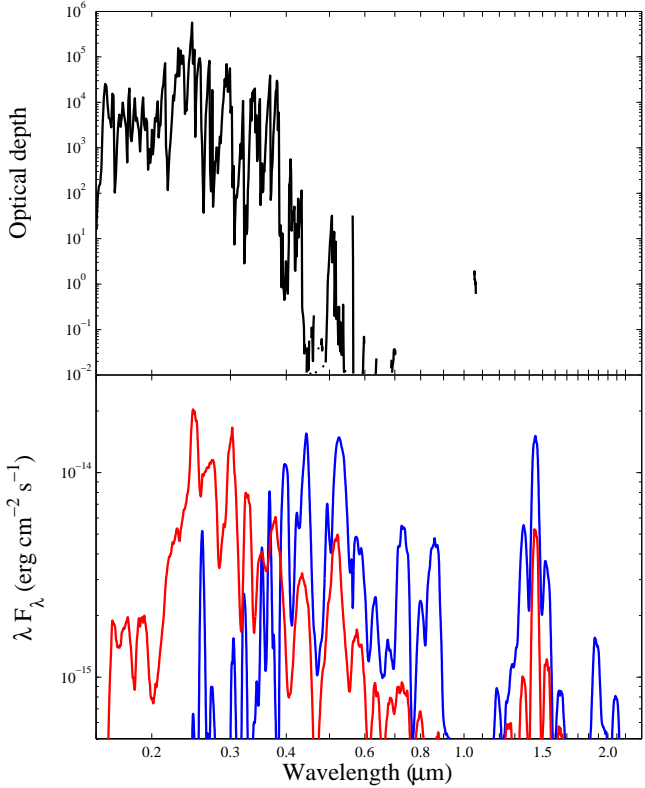}}
\caption{Lower panel: Comparison of  the energy distribution, $\lambda F_\lambda$,  of W7 with (blue) and without (red) multiple scattering, both calculated with the JFK11+ code. 
Upper panel: Line optical depth of the ejecta.
}
\label{figx1} 
\end{center}
\end{figure}

To demonstrate the importance of the radiative transfer treatment and the UV to optical conversion we show in Fig. \ref{figx1} the W7 spectrum from UV to NIR, with and without  non-local radiative transfer using the JFK11+ code.
 Without non-local radiative transfer we see  extremely strong UV lines below $\sim 4000$ \AA, mainly resulting from non-thermal ionizations to \ion{Fe}{2} and \ion{Fe}{3}, followed by recombinations in the UV, as well as excitations of \ion{Fe}{1}{-\small II}. In particular,  $\sim 1/3$ of all \ion{Fe}{1} excitations occur in a single transition, at $2348.3$ \AA,  followed mainly by the emission of lines at 2999.5 \AA \ and  $1.443 \mu$m. The latter dominates the NIR range (Fig. \ref{figx1})  and the non-thermal scenario may be tested by searching for this  signature in SN 2014J. The optical/NIR luminosity is  low, as most non-thermal excitation and ionizations lead to UV emission.

Including  non-local scattering and fluorescence  
the forest of optically thick lines (upper panel of Fig. \ref{figx1} ) decreases the UV flux below 4000 \AA \   by a factor 5.7, while the flux between 4000 - 10000 \AA \ and between $1 -2.5 \mu$m   increases by factors 3.4 and 3.2, respectively. Both multiple scatterings and fluorescence contribute to transferring the UV emissivity to optical and NIR flux. For example,    [\ion{Ca}{2}] 7300 \AA \ and the  \ion{Ca}{2} triplet are both powered by fluorescence following absorptions in the H and K lines.

\begin{figure}[!t]
\begin{center}
\resizebox{\hsize}{!}{\includegraphics{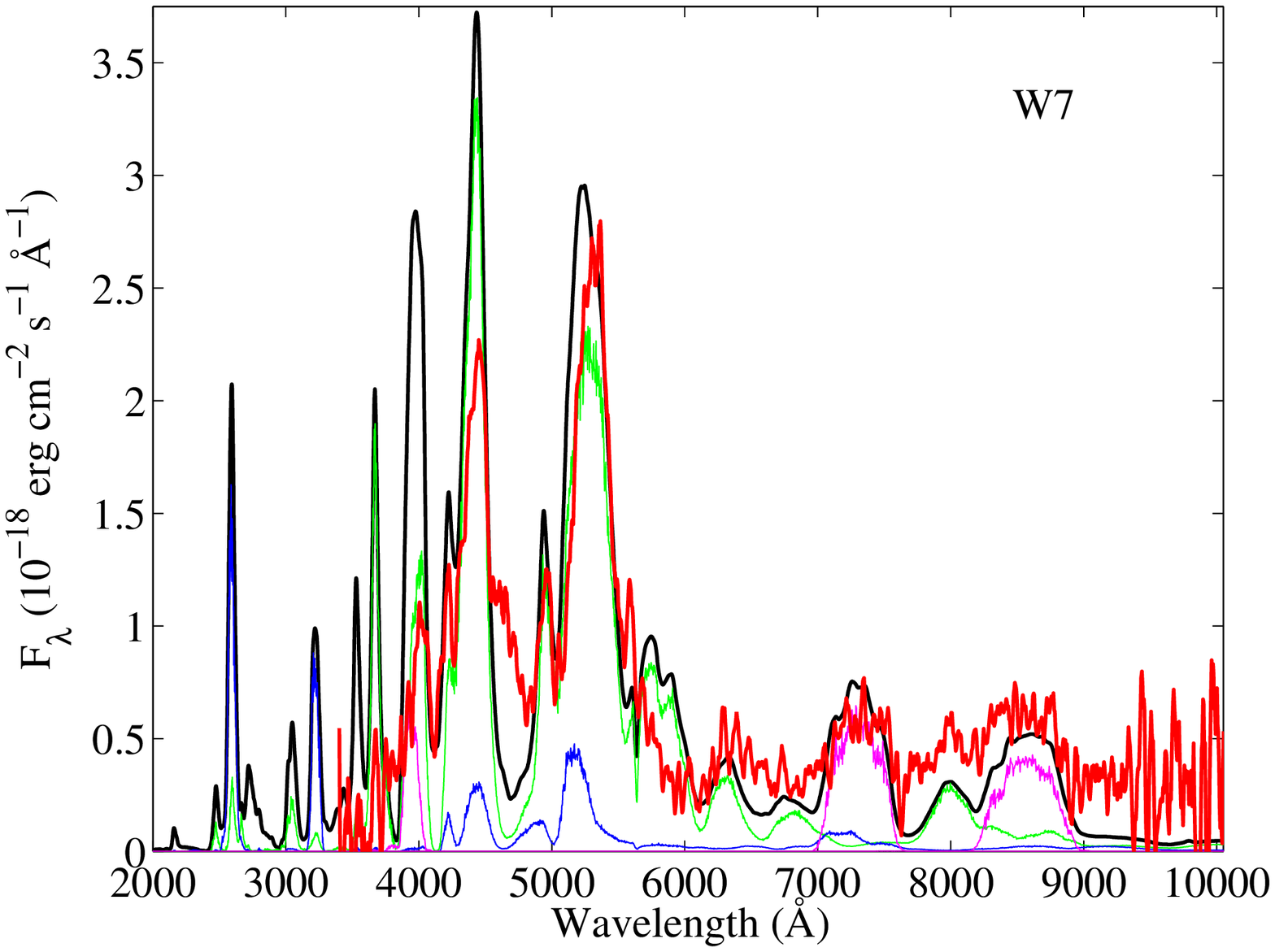}}
\resizebox{\hsize}{!}{\includegraphics{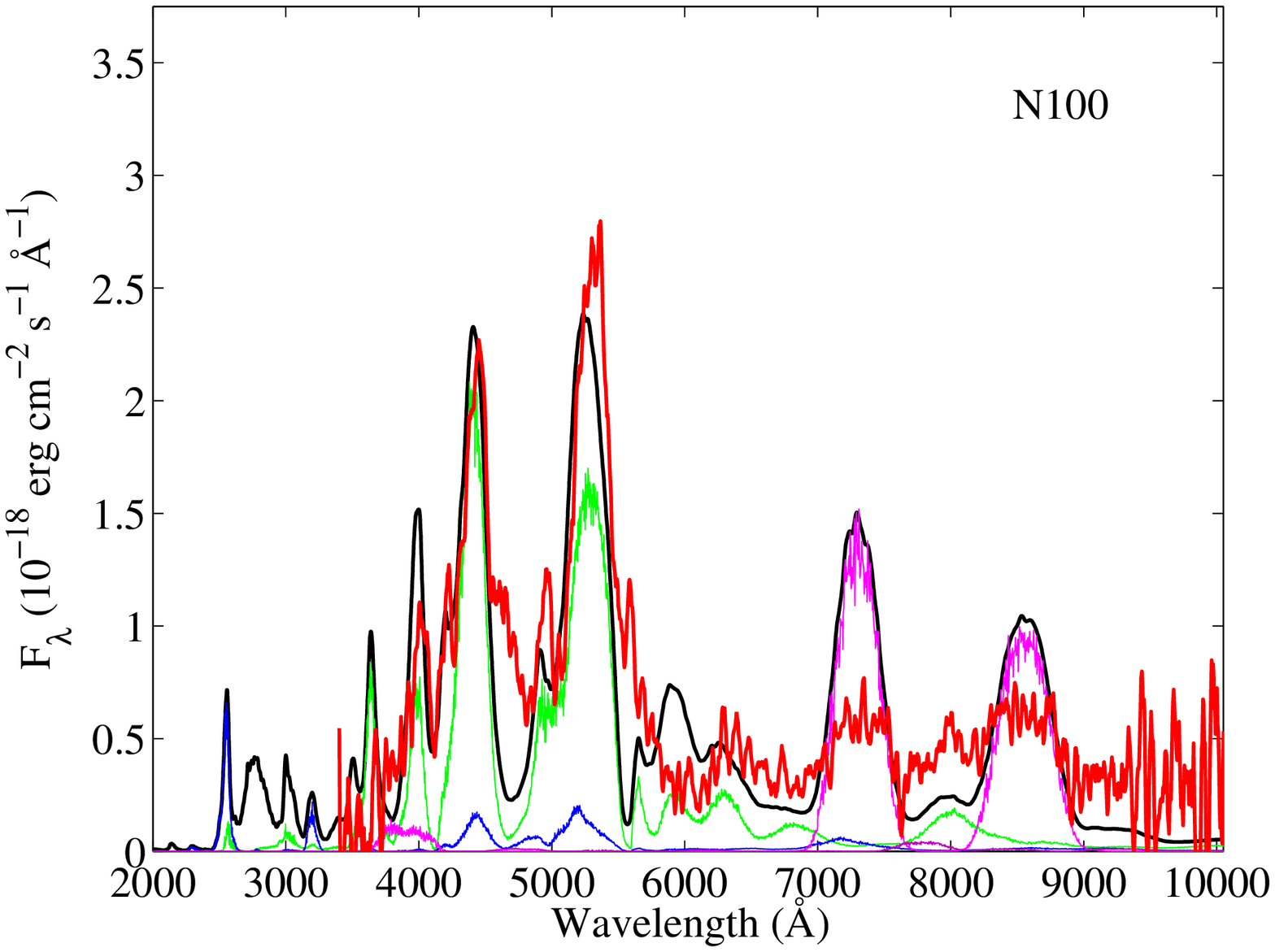}}
\caption{Optical spectrum of SN 2011fe at 1034 days from T2015 (red) together with the JFK11+ model spectrum (black) from the W7 (upper panel) and the N100 delayed detonation mode (lower panel). The contributions of \ion{Fe}{1} (green), \ion{Fe}{2} (blue) and \ion{Ca}{2} (magenta) are also shown.  
Both spectra are multiplied by a factor 2.0.}
\label{fig2}
\end{center}
\end{figure}

 In Fig. \ref{fig2} we  show the observed spectrum of SN 2011fe together with the W7 and N100 spectra  at 1034d calculated with the JFK11+ code. The models have been multiplied by a factor of 2,
a correction still needed to match the optical luminosity. As we discuss later, this can plausibly
be attributed to some combination of too low ${}^{57}$Ni mass
in the W7 model and lack of freeze-out in the JFK+11 code.

The spectrum is dominated by \ion{Fe}{1} lines as well as
[\ion{Ca}{2}] $\wll 7291, 7324$  and the Ca IR triplet. 
 In T2015 the feature at $\sim 6300$ \AA \ was interpreted as either [\ion{O}{1}] $\wll 6300, 6364$ or [\ion{Fe}{1}] $\wll 6359, 6231, 6394$.  In our synthetic spectra of both W7 and N100 it is dominated by \ion{Fe}{1}, with some contribution of \ion{Si}{1}{-\small II} in N100. There is therefore no need for any low velocity oxygen.  The feature  at $\sim 5900$ \AA \  is a mix of Na D and Fe I in N100. Unfortunately, Na is missing in the unburned high velocity region in W7, although already Fe I alone  tends to overproduce this.

 Calculating synthetic photometry for these spectra we find that the g and r band discrepancies between the photometry from the T2015 spectrum and the W7 model decrease from 3.0 mag and 2.2 mag  in the KF98 models to 0.36 mag and 0.89 mag, respectively, in the JFK11+ model (not including the factor 2 scaling in Fig. \ref{fig2})
 We note that the g-band contains most of the optical flux.  Given the uncertainties discussed below, we thus find a resolution to the long-standing problem of the IRC  not being observed in Type Ia supernovae: 
The omission of non-local radiative transfer effects in previous models seriously underestimates the optical/NIR flux at $t  \ga 500$ days.

We also note that the features at $\sim 4300$ \AA \ and $\sim 5200$ \AA \ are well reproduced in wavelength, and there is no need for any  ejecta asymmetry, as was suggested in T2015 from a comparison with the 300 day spectrum. This is a consequence of the shift from a \ion{Fe}{2}{-\small III} dominated spectrum to one dominated by \ion{Fe}{1}. 

There are many similarities of the 1000 days spectrum here to the conditions at $\sim 8$ years in the Fe-core  of SN 1987A, discussed by \cite{Jerkstrand2011}.  The densities are similar, 
$\sim 10^4 \ccm$, and the dominant excitation and ionization is by  leptons from the radioactive decays. Most of this energy is deposited in the iron rich regions in the core, resulting in \ion{Fe}{1} dominated spectra. 
From \citet[][Fig. 6 ]{Jerkstrand2011} one can  see that the  emissions following non-thermal excitations and ionizations below $\sim 3000$ \AA \ escapes in line gaps longwards of $\sim 4000$ \AA. 

Because of the low temperature, thermal, collisional processes are not important at this epoch, with the  exception of the mid/far-IR fine-structure lines, which account for $\sim 50 - 80 \%$ of the energy output (see below). This energy emerges mainly in [\ion{Fe}{2}] $25.99 \mu$m, and weaker [\ion{Fe}{3}] $ 22.93 \mu$m, [\ion{Fe}{1}] $ 24.04 \mu$m,  [\ion{Si}{2}] $ 34.81\mu$m and [\ion{Fe}{2}] $35.35 \mu$m. This part of the spectrum is unaffected by scattering. 

The main discrepancy in our modeling are the general level of the optical flux, which is a factor of $\sim 2$  fainter than observed.
In N100 there are also significant  differences in the  line ratios, in particular the ratio of the \ion{Ca}{2} lines to the \ion{Fe}{1} 4300, 5200 \AA \ multiplets.   In the W7 model there are also too strong line features at $\sim$ 2600, $\sim 3700$ \AA \ and at $\sim 4000$ \AA. There are, however, several factors which may contribute to these shortcomings. 

 An  important factor  is the quality  of atomic data, especially for \ion{Fe}{1}, where  most excitation  cross sections by non-thermal electrons are missing. Instead,  we use the Bethe approximation, which is  reasonable at high energy for permitted lines, but less accurate at low energies.
Fortunately, most excitations go to  permitted transitions. 

Recombinations to both \ion{Fe}{1} and \ion{Fe}{2} are 
responsible for a comparable flux to the excitations. Rates have been calculated for a large number of specific levels \citep{Nahar1997}, but half of the total rate is to higher levels,  and is in our model atom distributed among the highest levels according to statistical weights, introducing an uncertainty in individual lines.  

Uncertainty in the explosive burning conditions is associated with at least a factor 2 uncertainty in
the ${}^{57}$Ni mass, which
accounts for$\ga 50\%$ of the energy input at this epoch. The density distribution at high velocity also shows a large  variation between different models. An underestimate of the 'Fe' abundances and/or the density at high velocities results in an underestimate of the UV-scattering, which may
be the reason for the model overproduction of lines below $\sim 4000$ \AA. 

The most important factor is probably the steady state assumption in the JFK11+ models. Because of the low  density  the balance between heating and cooling, as well as between ionization and recombination, may break down, usually known as 'freeze-out'  \citep{Fransson1993}.  From a comparison of a steady-state and a fully time-dependent calculation of the W7 model with the KF98 code, we find that later than $\sim 700$ days  time dependent effects become increasingly important (Fig. \ref{fig1b}). 
 This leads to an underestimate of the flux  by  0.35 and 0.86 mags in the V and R bands respectively at 1000 days  in the steady-state compared to the time dependent case.  We note the larger difference in the R-band, as is also seen for the JFK11+ model  when compared to observations.

Taking these uncertainties
and especially the freeze-out,  
into account, the model provides a satisfactory reproduction of the main properties of the 1034 day spectrum and photometry.

\begin{figure}[!t]
\begin{center}
\resizebox{\hsize}{!}{\includegraphics{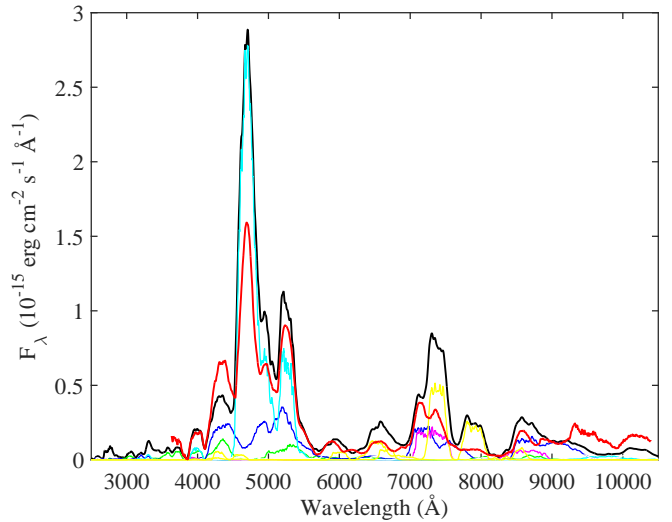}}
\caption{Optical spectrum of SN 2011fe at 331 days from T2015 (red) together with the  JFK11+ model spectrum from the W7 model (black).  Individual contributions are the same as in Fig. \ref{fig2} with the addition of \ion{Fe}{3} (cyan) and \ion{Ni}{2}{-\scriptsize III} (yellow). 
 }
\label{fig0}
\end{center}
\end{figure}

As a further test we show in Fig. \ref{fig0} the JFK11+ model for W7 at 331d, compared with observations from T2015. 
In general, there is agreement between the model and the strongest features. The optical spectrum at this epoch is mainly dominated by blends of [\ion{Fe}{2}{-\small III}] and weaker features of \ion{Ca}{2} and [\ion{Ni}{2}].  The overproduction of the [\ion{Ni}{2}{-\small III}] lines between 6000-8200 \AA \ indicates that W7 contains too much stable Ni. The analysis by \cite{Mazzali2015} also favoured relatively small amounts of stable nickel in the ejecta.  
The too strong \ion{Fe}{3} multiplet at $\sim 4700$ \AA \  suggests a too high ionization of the ejecta.  Despite these shortcomings the model can  qualitatively reproduce the spectrum {showing that one and the same model works at both 300d and 1000d.  
}

\section{Discussion}
\label{sec-disc}

After the  IRC, the temperature is far too low for thermal excitations of anything except for the mid/far-IR fine structure lines. Because of the concomitant decrease in the ionization
an increasing fraction of the energy goes into non-thermal excitations and ionizations rather than heating \citep{Kozma1992}.  
Recombinations, following the non-thermal ionizations to \ion{Fe}{2}, are mainly to  high-excitation levels of \ion{Fe}{1}. This is also the case for the excitations, where the high-lying permitted transitions have the largest cross sections for the non-thermal electrons and positrons. Therefore both these processes result in populations of primarily high levels, which  de-excite by permitted transitions in the UV. 

\begin{figure}[!t!]
\begin{center}
\resizebox{\hsize}{!}{\includegraphics[angle=0]{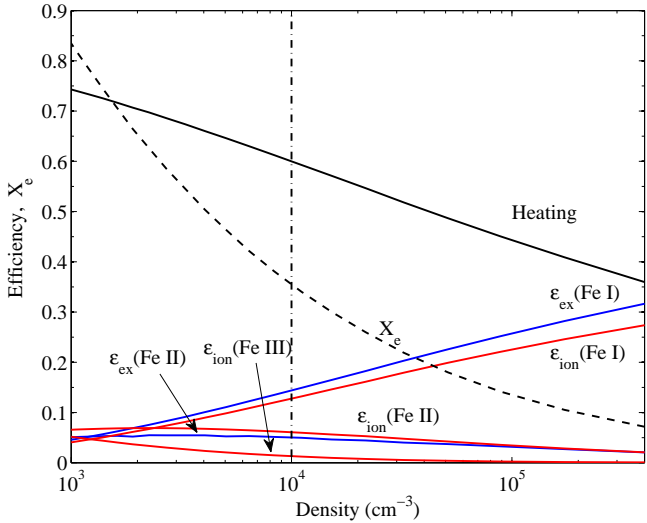}}
\caption{Electron fraction $X_e$ (black, dashed) and energy fractions going into excitations (blue), ionizations (red) and heating (black, solid) by non-thermal electrons and positrons for a pure iron plasma at 1000d.  The vertical dash-dotted line is the density at 1000 days for  a uniform sphere of $1.4 \Msun$ expanding at $10^4 \kms$ .}
\label{fig17}
\end{center}
\end{figure}

In Fig. \ref{fig17} we show the change of  ionization with density at 1000 days for a  model where we calculate the state of ionization and energy deposition for a pure iron plasma, but ignore all radiative transfer effects and only include  cooling by the [\ion{Fe}{1}{-\small II}] fine-structure lines. Because photoionizations are not included the degree of ionization may be somewhat underestimated, but the qualitative features should be correct. We assume M(${}^{56}$Ni) $= 0.6 \Msun$ and M(${}^{57}$Ni)$= 2 \times 10^{-2} \Msun$ and constant energy input per unit mass. Gamma-rays are  neglected, but make only a minor contribution. 

From this we see that in the  range $1 \times 10^3 - 4 \times 10^5 \ccm$ the electron fraction, $X_{\rm e}$, varies from $\sim 1.0$ to  $\sim  0.07$.
Because the densities below $10^4 \kms$ vary between $\sim 5\times 10^3 \ccm$ to $\sim 5\times 10^4 \ccm$
 in the explosion models (for no clumping),
 the dominance of \ion{Fe}{1} features should be a robust result at these epochs. There is simply not enough radioactive power to maintain a higher ionization state.  Between $1 \times 10^3 - 4 \times 10^5 \ccm$  the heating efficiency varies between $\epsilon_{\rm h} \approx 0.4-0.8$. Because the local ionization is roughly $\propto (X_{56+57} / \rho)^{1/2}$, where $X_{56+57}$ is the initial abundance of the radioactive ${}^{56}$Ni and ${}^{57}$Ni isotopes, $\epsilon_{\rm h} $ depends on the density where the radioactive isotopes are abundant, which differs between different explosion models.  With the JFK11+ code the deposition weighted heating efficiencies are $\epsilon_{\rm h} \approx 0.76$ for W7 and $\epsilon_{\rm h} \approx 0.73$ for N100. 

Because  the fraction going into heating is emitted as mid/far-IR radiation, while the rest is converted to optical/NIR emission, 
an increase in the ionization and excitation efficiency with density  can have a strong effect on the optical flux. 
Clumping could increase the optical flux of our models and 
help alleviate the discrepancy with the observations  \citep[][]{Kozma2005,Leloudas2009}.

\section{Conclusions}
\label{sec-conculsions}

In this Letter we have studied the importance of  radiative transfer  in the nebular phase  of Type Ia SNe.
 Without non-local radiative transfer, only $\sim 5\%$ of the deposited energy emerges in the optical/NIR
at 1000d, the rest being in the MIR ($\sim 80\%$) and UV ($\sim 15 \%$). Non-local scattering and fluoresence, however, converts most of the UV flux into the optical/NIR, raising this by a factor $\sim 4$ to $\sim 20 \%$.
 Because of the dominance of non-thermal processes and the UV to optical conversion  there is therefore no contradiction between the drop in temperature 
caused by the IRC and a sustained optical flux roughly consistent with the radioactive decay. Between $50-80 \%$ of the non-thermal deposition still goes to heating leading to mid-IR line cooling, in particular by [\ion{Fe}{2}] $\wl 25.99 \mu$m. With this model improvement, we find a good general reproduction of the spectrum of SN 2011fe at 1000 days with a spectrum dominated by \ion{Fe}{1},  
which explains  the 'shift' of some lines at this epoch compared to the thermally dominated \ion{Fe}{2}{-\small III} spectrum at 300 days, discussed by T2015. 

We also find  a strong need of radioactive input from ${}^{57}$Co at $\sim 1000$ days. Without this the optical spectrum would be underproduced by a factor $\sim 4$, compared to the factor $\sim 2$ in our models, which include $\sim 2 \times 10^{-2} \msun$ of ${}^{57}$Ni. A ${}^{57}$Ni input at this level together with the freeze-out effects appears to be necessary to give the observed  optical flux level at this late epoch.

\acknowledgments

We are grateful to Markus Kromer, Stefan  Taubenberger,  Melissa Graham and Ken Nomoto for discussions and for models and to the referee for very useful comments. This research was supported by the Swedish Research Council and
National Space Board.


\begin{thebibliography}{22}
\expandafter\ifx\csname natexlab\endcsname\relax\def\natexlab#1{#1}\fi

\bibitem[{{Axelrod}(1980)}]{Axelrod1980}
{Axelrod}, T.~S. 1980, PhD thesis, California Univ., Santa Cruz.

\bibitem[{{Fransson} {et~al.}(1996){Fransson}, {Houck}, \&
  {Kozma}}]{Fransson1996}
{Fransson}, C., {Houck}, J., \& {Kozma}, C. 1996, in IAU Colloq. 145:
  Supernovae and Supernova Remnants, ed. T.~S. {Kuhn}, 211

\bibitem[{{Fransson} \& {Kozma}(1993)}]{Fransson1993}
{Fransson}, C., \& {Kozma}, C. 1993, \apjl, 408, L25

\bibitem[Graham et al.(2015)]{Graham2015} Graham, M.~L., Nugent, 
P.~E., Sullivan, M., et al.\ 2015, arXiv:1502.00646 

\bibitem[{{Graur} {et~al.}(2015){Graur}, {Zurek}, {Shara}, \&
  {Riess}}]{Graur2015}
{Graur}, O., {Zurek}, D., {Shara}, M.~M., \& {Riess}, A.~G. 2015, ArXiv
  e-prints 1505.00777

\bibitem[{{Iwamoto} {et~al.}(1999){Iwamoto}, {Brachwitz}, {Nomoto},
  {Kishimoto}, {Umeda}, {Hix}, \& {Thielemann}}]{Iwamoto1999}
{Iwamoto}, K., {Brachwitz}, F., {Nomoto}, K., {et~al.} 1999, \apjs, 125, 439

\bibitem[{{Jerkstrand} {et~al.}(2015){Jerkstrand}, {Ergon}, {Smartt},
  {Fransson}, {Sollerman}, {Taubenberger}, {Bersten}, \&
  {Spyromilio}}]{Jerkstrand2015}
{Jerkstrand}, A., {Ergon}, M., {Smartt}, S.~J., {et~al.} 2015, \aap, 573, A12

\bibitem[{{Jerkstrand} {et~al.}(2011){Jerkstrand}, {Fransson}, \&
  {Kozma}}]{Jerkstrand2011}
{Jerkstrand}, A., {Fransson}, C., \& {Kozma}, C. 2011, \aap, 530, A45

\bibitem[{{Kerzendorf} {et~al.}(2014){Kerzendorf}, {Taubenberger},
  {Seitenzahl}, \& {Ruiter}}]{Kerzendorf2014}
{Kerzendorf}, W.~E., {Taubenberger}, S., {Seitenzahl}, I.~R., \& {Ruiter},
  A.~J. 2014, \apjl, 796, L26

\bibitem[{{Kozma} \& {Fransson}(1992)}]{Kozma1992}
{Kozma}, C., \& {Fransson}, C. 1992, \apj, 390, 602

\bibitem[Kozma 
\& Fransson(1998a)]{Kozma1998a} Kozma, C., \& Fransson, C.\ 1998, \apj, 496, 946 

\bibitem[{{Kozma} \& {Fransson}(1998b)}]{Kozma1998b}
---. 1998, \apj, 497, 431

\bibitem[{{Kozma} {et~al.}(2005){Kozma}, {Fransson}, {Hillebrandt},
  {Travaglio}, {Sollerman}, {Reinecke}, {R{\"o}pke}, \&
  {Spyromilio}}]{Kozma2005}
{Kozma}, C., {Fransson}, C., {Hillebrandt}, W., {et~al.} 2005, \aap, 437, 983

\bibitem[{{Leloudas} {et~al.}(2009){Leloudas}, {Stritzinger}, {Sollerman},
  {Burns}, {Kozma}, {Krisciunas}, {Maund}, {Milne}, {Filippenko}, {Fransson},
  {Ganeshalingam}, {Hamuy}, {Li}, {Phillips}, {Schmidt}, {Skottfelt},
  {Taubenberger}, {Boldt}, {Fynbo}, {Gonzalez}, {Salvo}, \&
  {Thomas-Osip}}]{Leloudas2009}
{Leloudas}, G., {Stritzinger}, M.~D., {Sollerman}, J., {et~al.} 2009, \aap,
  505, 265
  
\bibitem[Mazzali et al.(2015)]{Mazzali2015} Mazzali, P.~A., 
Sullivan, M., Filippenko, A.~V., et al.\ 2015, \mnras, 450, 2631 

\bibitem[{{Nahar} {et~al.}(1997){Nahar}, {Bautista}, \& {Pradhan}}]{Nahar1997}
{Nahar}, S.~N., {Bautista}, M.~A., \& {Pradhan}, A.~K. 1997, \apj, 479, 497

\bibitem[{{Nomoto} {et~al.}(1984){Nomoto}, {Thielemann}, \&
  {Yokoi}}]{Nomoto1984}
{Nomoto}, K., {Thielemann}, F.-K., \& {Yokoi}, K. 1984, \apj, 286, 644

\bibitem[Nugent et al.(2011)]{Nugent2011} Nugent, P.~E., Sullivan, 
M., Cenko, S.~B., et al.\ 2011, \nat, 480, 344 

\bibitem[{{Pereira} {et~al.}(2013){Pereira}, {Thomas}, {Aldering}, {Antilogus},
  {Baltay}, {Benitez-Herrera}, {Bongard}, {Buton}, {Canto}, {Cellier-Holzem},
  {Chen}, {Childress}, {Chotard}, {Copin}, {Fakhouri}, {Fink}, {Fouchez},
  {Gangler}, {Guy}, {Hillebrandt}, {Hsiao}, {Kerschhaggl}, {Kowalski},
  {Kromer}, {Nordin}, {Nugent}, {Paech}, {Pain}, {P{\'e}contal}, {Perlmutter},
  {Rabinowitz}, {Rigault}, {Runge}, {Saunders}, {Smadja}, {Tao},
  {Taubenberger}, {Tilquin}, \& {Wu}}]{Pereira2013}
{Pereira}, R., {Thomas}, R.~C., {Aldering}, G., {et~al.} 2013, \aap, 554, A27

\bibitem[R{\"o}pke et al.(2012)]{Ropke2012} R{\"o}pke, F.~K., 
Kromer, M., Seitenzahl, I.~R., et al.\ 2012, \apjl, 750, L19 

\bibitem[{{Seitenzahl} {et~al.}(2009){Seitenzahl}, {Taubenberger}, \&
  {Sim}}]{Seitenzahl2009}
{Seitenzahl}, I.~R., {Taubenberger}, S., \& {Sim}, S.~A. 2009, \mnras, 400, 531

\bibitem[{{Seitenzahl} {et~al.}(2013){Seitenzahl}, {Ciaraldi-Schoolmann},
  {R{\"o}pke}, {Fink}, {Hillebrandt}, {Kromer}, {Pakmor}, {Ruiter}, {Sim}, \&
  {Taubenberger}}]{Seitenzahl2013}
{Seitenzahl}, I.~R., {Ciaraldi-Schoolmann}, F., {R{\"o}pke}, F.~K., {et~al.}
  2013, \mnras, 429, 1156

\bibitem[{{Shappee} \& {Stanek}(2011)}]{Shappee2011}
{Shappee}, B.~J., \& {Stanek}, K.~Z. 2011, \apj, 733, 124

\bibitem[{{Sollerman} {et~al.}(2004){Sollerman}, {Lindahl}, {Kozma}, {Challis},
  {Filippenko}, {Fransson}, {Garnavich}, {Leibundgut}, {Li}, {Lundqvist},
  {Milne}, {Spyromilio}, \& {Kirshner}}]{Sollerman2004}
{Sollerman}, J., {Lindahl}, J., {Kozma}, C., {et~al.} 2004, \aap, 428, 555

\bibitem[{{Taubenberger} {et~al.}(2015){Taubenberger}, {Elias-Rosa},
  {Kerzendorf}, {Hachinger}, {Spyromilio}, {Fransson}, {Kromer}, {Ruiter},
  {Seitenzahl}, {Benetti}, {Cappellaro}, {Pastorello}, {Turatto}, \&
  {Marchetti}}]{Taubenberger2015}
{Taubenberger}, S., {Elias-Rosa}, N., {Kerzendorf}, W.~E., {et~al.} 2015,
  \mnras, 448, L48 (T2015)

\bibitem[{{Tsvetkov} {et~al.}(2013){Tsvetkov}, {Shugarov}, {Volkov},
  {Goranskij}, {Pavlyuk}, {Katysheva}, {Barsukova}, \& {Valeev}}]{Tsvetkov2013}
{Tsvetkov}, D.~Y., {Shugarov}, S.~Y., {Volkov}, I.~M., {et~al.} 2013,
  Contributions of the Astronomical Observatory Skalnate Pleso, 43, 94

\end{thebibliography}
\end{document}